\documentstyle[seceq,epsf,wrapft,twoside,mbf]{ptptex}
\notypesetlogo  %comment in if to eliminate PTPTeX logo
\title{Quark level linear ${\mib \sigma}$ model (L${\mib{\sigma}}$M) via loop graphs
}
\author{%
M.D. {\sc Scadron}}
\inst{%
 Physics Department, University of Arizona, Tucson, AZ, 85721 USA
}
\abst{%
First, the L$\sigma$M is nonperturbatively solved via loop-order gap equations.
Then the Nambu-Goldstone theorem (NGT) is expressed in L$\sigma$M language. 
Next, the Lee null tadpole sum for this SU(2) L$\sigma$M theory is shown 
to require $N_c$ = 3. Then L$\sigma$M s-wave chiral cancellations are studied 
in tree-and in loop-order. Also, vector meson dominance (VMD) is shown 
to follow from this L$\sigma$M. Next, the scalar nonet is studied in the 
infinite momentum frame (IMF). Finally, the L$\sigma$M is suggested 
as the infrared limit of nonperturbative QCD.
\begin{center}
{\scriptsize {\it Invited Talk at the scalar meson workshop, Kyoto, June 2000}}
\end{center}
}
\begin{document}
\maketitle

\setcounter{tocdepth}{4}

\section{Introduction}

To begin, we give the original\cite{rf1} tree-level chiral-broken
SU(2) interacting L$\sigma$M lagrangian density, but after the  
spontaneous symmetry breaking (SSB) shift:
\begin{eqnarray}
 {\cal L}_{L\sigma M}^{int}=g\bar{\Psi}(\sigma'+i\gamma_5{\mbf \tau}\cdot{\mbf \pi})\Psi 
   + g'\sigma'(\sigma'^2+{\mbf \pi}^2)
 -\lambda(\sigma'^2+{\mbf \pi}^2)^2/4.
\label{eq1}
\end{eqnarray}
In refs.\citen{rf1} the couplings $g$, $g'$, $\lambda$ in (\ref{eq1}) 
satisfy the quark-level Goldberger-Treiman relation (GTR) for 
$f_\pi\approx$ 93 MeV 
and $f_\pi\sim$ 90MeV in the chiral limit (CL):
\begin{eqnarray}
 g=m_q/f_\pi, \ \ \ g'=m_\sigma^2/2f_\pi = \lambda f_\pi.
\label{eq2}
\end{eqnarray}

We work in loop order and dynamically generate mass terms in 
(\ref{eq1}) via nonperturbative Nambu-type gap equations 
$\delta f_\pi = f_\pi$, $\delta m_q = m_q$.
The CL $m_\pi$ = 0, corresponds to $<0|\partial A|\pi> = 0$ 
for $<0|A_\mu^3|\pi^0> = if_\pi q_\mu$. 
The latter requires the GTR $m_q = f_\pi g$ to be valid 
in tree and loop order, fixing $g$, $g'$, $\lambda$ in loop order. 

In \S \ref{sec2}, \ref{sec3}, this quark-level L$\sigma$M is 
nonperturbatively solved 
via loop-order gap equations. In \S \ref{sec4}, the NGT is expressed 
in L$\sigma$M language with charge radius $r_\pi = 1/m_q$ characterizing 
quark fusion for the tightly bound $q\bar{q}$ pion. 
In \S \ref{sec5}, the Lee null tadpole sum is shown to require 
$N_c = 3$ for the true vacuum. \S \ref{sec6} discusses $s$-wave chiral 
cancellations in the L$\sigma$M. 
\S \ref{sec7} shows VMD follows directly from the L$\sigma$M. 
\S \ref{sec8} studies the scalar meson nonet in the IMF. 
Finally \S \ref{sec9} suggests this L$\sigma$M is the infrared limit of 
nonperturbative QCD. We give our conclusions in \S \ref{sec10}.

\section{Quark loop gap equations}
\label{sec2}

\begin{figure}[t]
  \epsfysize=1.5 cm
 \centerline{\epsffile{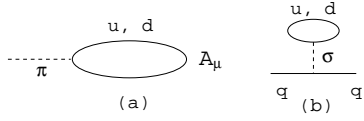}}
 \caption{Quark loop for $f_\pi$(a), quark tadpole loop for $m_q$(b)}
  \label{fig1}
\end{figure}
First we compute $\delta f_\pi= f_\pi$ in the CL via the $u$ and $d$ quark 
loops of Fig.\ref{fig1}a.
Replacing $f_\pi$ by $m_q/g$ and taking the quark trace, giving $4m_q q_\mu$, 
the factors $m_q q_\mu$ cancel, requiring the CL log-divergent gap equation 
(LDGE)\cite{rf2,rf3}, $\bar{\mbf d}^4 p = d^4p /(2\pi)^4$:
\begin{eqnarray}
 1=-4iN_c g^2\int (p^2-m_q^2)^{-2} \bar{\mbf d}^4 p.
\label{eq3}
\end{eqnarray}

Anticipating $g\sim$ 320 MeV/90MeV $\sim$3.6 from the CL GTR, this LDGE (\ref{eq3}) 
suggests an UV cutoff $\Lambda\sim$750 MeV. Such a 
750 MeV cutoff separates L$\sigma$M elementary particle $\sigma$(600) 
$< \Lambda$ from bound states $\rho$(770), $\omega$(780), $a_1$(1260)$ > \Lambda$. 
This is a $Z$ = 0 compositeness condition\cite{rf4}, requiring 
$g = 2\pi/\sqrt{N_c}$. We later derive this from 
our dynamical symmetry breaking (DSB) loop order L$\sigma$M. 

Next we study $\delta m_q = m_q$ in the CL, with zero current quark mass; 
$m_q$ is the nonstrange constituent 
quark mass. The needed mass gap is formed via the quadratically divergent 
quark tadpole loop of Fig.\ref{fig1}b; additional quark $\pi$- and 
$\sigma$-mediated self-energy graphs then cancel\cite{rf3}, 
giving the quadratic divergent mass gap
\begin{eqnarray}
 1={8iN_c g^2}/( -m_\sigma^2)\cdot \int(p^2-m_q^2)^{-1} \bar{\mbf d}^4 p.
\label{eq4}
\end{eqnarray}
Here the $q^2$ = 0 tadpole $\sigma$ propagator $(0 - m^2_\sigma)^{-1}$ means 
the right-hand side (rhs) of the integral in eq.(\ref{eq4}) acts as a 
counterterm quadratic divergent NJL\cite{rf5} mass gap.

References \citen{rf3} first subtract the quadratic-from the log-divergent 
integrals of eqs. (\ref{eq3}), (\ref{eq4}) to form the dimensional 
regularization (dim. reg.) lemma for $2l=4$: 
\begin{eqnarray}
 \int\bar{\mbf d}^4 p \left[\frac{m_q^2}{(p^2-m_q^2)^2}-\frac{1}{p^2-m_q^2}\right] =\lim_{l\rightarrow 2}
 \frac{im_q^{2l-2}}{(4\pi)^l} \left[\Gamma(2-l)+\Gamma(1-l)\right]
=\frac{-im_q^2}{(4\pi)^2}.\ \ \ \ \ \ \ \
\label{eq5}
\end{eqnarray}
This dim. reg. lemma (\ref{eq5}) follows because 
$\Gamma(2-l) + \Gamma(1-l) \rightarrow -1$ as $l\rightarrow 2$ due 
to the gamma function defining identity $\Gamma(z + 1) = z\Gamma(z)$. 
This lemma eq.(\ref{eq5}) is more general than dim. reg.; (i) use partial 
fractions to write
\begin{eqnarray}
 \frac{m^2}{(p^2-m^2)^2}- \frac{1}{p^2-m^2}
 =\frac{1}{p^2}\left[\frac{m^4}{(p^2-m^2)^2}-1\right],
\label{eq6}
\end{eqnarray}
(ii) integrate eq.(\ref{eq6}) via $\bar{\mbf d}^4 p$ and neglect the latter 
massless tadpole $\int\bar{\mbf d}^4 p/p^2 = 0$ 
(as is also done in dim. reg., analytic, zeta function and Pauli-Villars 
regularization\cite{rf3}) (iii) Wick rotate 
$d^4p = i \pi^2 p_E^2 dp_E^2$ in the integral over eq.(\ref{eq6}) 
to find 
\begin{eqnarray}
 \int \bar{\mbf d}^4 p 
\left[ \frac{m^2}{(p^2-m^2)^2} -\frac{1}{p^2-m^2} \right]
 =
  -\frac{im^4}{(4\pi)^2}\int_0^\infty \frac{ dp_E^2}{(p_E^2+m^2)^2}
 =\frac{-im^2}{(4\pi)^2}.\ \ \ 
\label{eq7}
\end{eqnarray}
So (\ref{eq7}) gives the dim.reg.lemma (\ref{eq5}); 
both are {\it regularization scheme independent}.

Following refs.\citen{rf3} we combine eqs. (\ref{eq5}) or (\ref{eq7}) 
with the LDGE (\ref{eq3}) to solve the quadratically divergent mass gap 
integral (\ref{eq4}) as 
\begin{eqnarray}
 m_\sigma^2=2m_q^2(1+g^2N_c/4\pi^2).
\label{eq8}
\end{eqnarray}
\begin{figure}[t]
\parbox{\halftext}{
  \epsfxsize=6. cm
  \epsfysize=1. cm
  \centerline{\epsffile{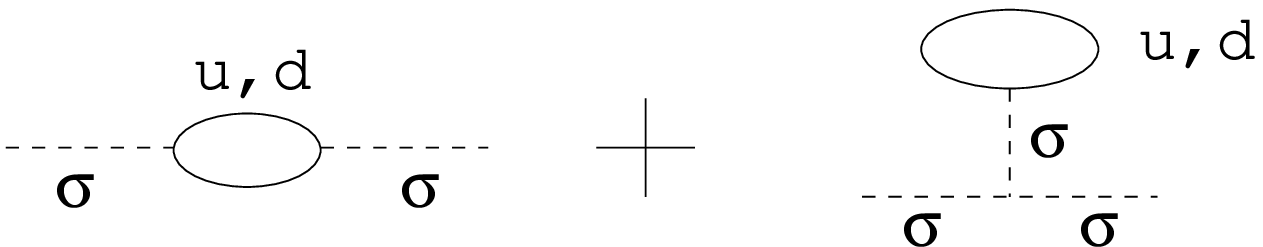}}
  \caption{Quark bubble plus quark tadpole loop for $m_\sigma^2$.}
  \label{fig2}}
 \hspace{4mm}
 \parbox{\halftext}{
  \epsfxsize=6. cm
  \epsfysize=1. cm
  \centerline{\epsffile{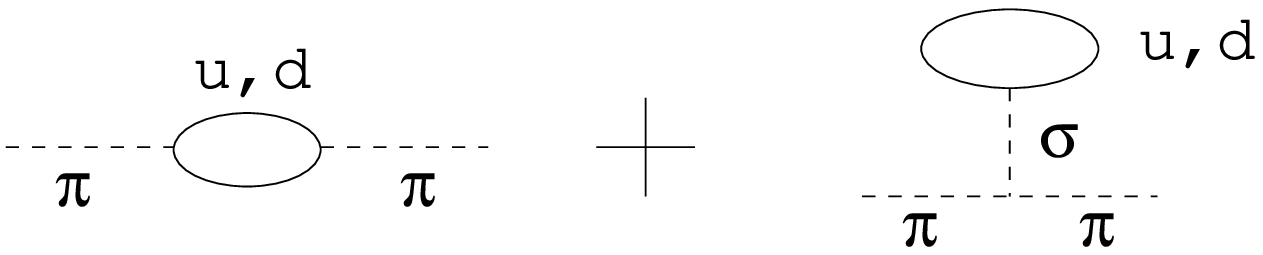}}
  \caption{Quark bubble plus quark tadpole loop for $m_\pi^2$.}
  \label{fig3}\vspace{0mm}}
\end{figure}
Also the Fig.\ref{fig2} quark bubble plus tadpole graphs dynamically generate 
the $\sigma$ mass\cite{rf3}:
\begin{eqnarray}
 m_\sigma^2=16iN_cg^2\int \bar{\mbf d}^4p \left[\frac{m^2_q}{(p^2-m^2_q)^2}
 -\frac{1}{p^2-m_q^2}\right]=\frac{N_cg^2m_q^2}{\pi^2},
\label{eq9}
\end{eqnarray}
where we have deduced the rhs of eq.(\ref{eq9}) by using (\ref{eq5}) or 
(\ref{eq7}). Finally solving the two equations (\ref{eq8}) and 
(\ref{eq9}) for the two unknowns $m^2_\sigma / m^2_q$ and $g^2 N_c$, 
one finds\cite{rf3} 
\begin{eqnarray}
 m_\sigma=2m_q,\ \ \ \ \ \ \ \ \ \ g=2\pi/\sqrt{N_c}.
\label{eq10}
\end{eqnarray}
Not surprisingly, the lhs equation in (\ref{eq10}) is the famous NJL four 
quark result\cite{rf5}, earlier anticipated for the L$\sigma$M in 
refs.\citen{rf6}. The rhs equation in (\ref{eq10}) is also 
the consequence of the Z=0 compositeness condition\cite{rf4}, as noted earlier.

Finally we compute $m^2_\pi$ from the analog pion bubble plus tadpole graphs 
of Fig.\ref{fig3}. Since both quark loops (ql) are 
quadratic divergent in the CL, one finds\cite{rf2,rf3}
\begin{eqnarray}
 m_{\pi, ql}^2=4iN_c[2g^2-4gg'm_q/m_\sigma^2]\int(p^2-m_q^2)^{-1} 
 \bar{\mbf d}^4p=0;
\ 
 g'=m_\sigma^2/2f_\pi,
\label{eq12}
\end{eqnarray}
using the GTR. Not suprisingly, eq.(\ref{eq12}) is the dynamical 
version of the SSB (\ref{eq2}).

\section{Loop order three- and four-point functions}
\label{sec3}

\begin{figure}[t]
\parbox{\halftext}{
  \epsfxsize=6. cm
  \epsfysize=.8 cm
  \centerline{\epsffile{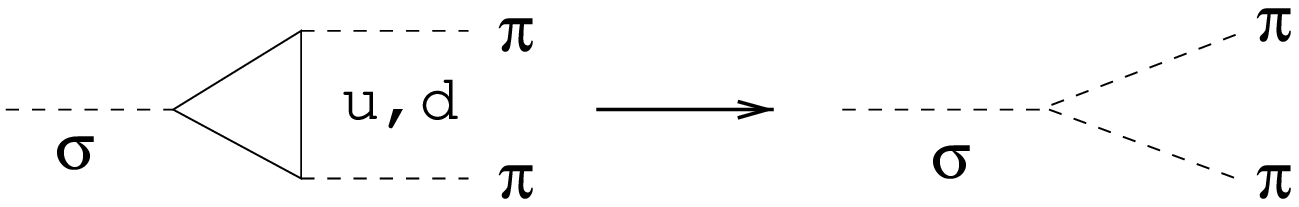}}
  \caption{Quark triangle shrinks to point for $m_\sigma\rightarrow\pi\pi$.}
  \label{fig4}}
 \hspace{4mm}
 \parbox{\halftext}{
  \epsfxsize=6. cm
  \epsfysize=.8 cm
  \centerline{\epsffile{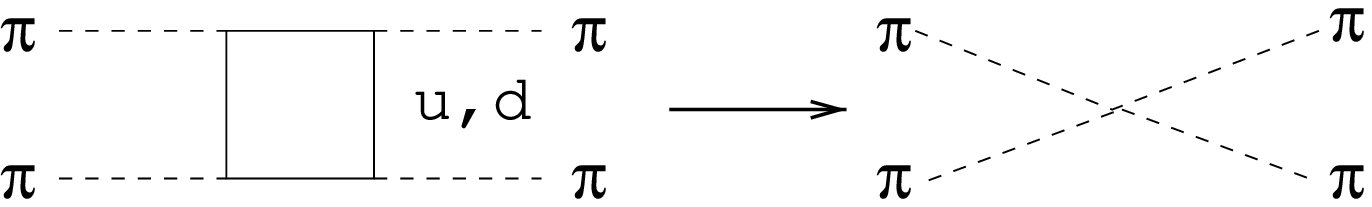}}
  \caption{Quark box shrinks to point contact for $\pi\pi\rightarrow\pi\pi$.}
  \label{fig5}\vspace{0mm}}
\end{figure}

Having studied all 2-point functions in \S \ref{sec2}, we now look at 
3- and 4-point functions. In the CL the $u$ and $d$ quark 
loops of Fig.\ref{fig4} generate $g_{\sigma\pi\pi}$\cite{rf2,rf3} as
\begin{eqnarray}
 g_{\sigma\pi\pi}=-8ig^3N_cm_q\int (p^2-m_q^2)^{-2} \bar{\mbf d}^4p =2gm_q
\label{eq13}
\end{eqnarray}
by virtue of the LDGE (\ref{eq3}). Using the GTR and 
$m_\sigma = 2m_q$, eq.(\ref{eq13}) reduces to
\begin{eqnarray}
 g_{\sigma\pi\pi}=2gm_q=m_\sigma^2/2f_\pi=g'.
\label{eq14}
\end{eqnarray}
In effect, the $g_{\sigma\pi\pi}$ loop of Fig.\ref{fig4} ``shrinks" to 
the L$\sigma$M cubic meson coupling $g'$ in the tree-level lagrangian 
eq.(\ref{eq1}), but only when $m_\sigma = 2m_q$ and $g/m_q = 1/f_\pi$.

Next we study the 4-point $\pi\pi$ quark box of Fig.\ref{fig5},
giving a CL log divergence\cite{rf3}:
\begin{eqnarray}
 \lambda_{box}=-8iN_cg^4\int (p^2-m_q^2)^{-2} \bar{\mbf d}^4p =2g^2
 = g'/f_\pi=\lambda_{tree},
 \label{eq15}
\end{eqnarray}
employing the LDGE (\ref{eq3}) to reduce (\ref{eq15}) to $2g^2$. 
Equation(\ref{eq15}) shrinks to
$\lambda_{tree}$, by virtue of eq.(\ref{eq2}). 
Substituting (\ref{eq10}) into (\ref{eq15}), 
we find $\lambda = 8\pi^2/N_c$.

We have dynamically generated the entire 
L$\sigma$M lagrangian (\ref{eq1}), but using the DSB true 
vacuum, satisfying specific values of $g$, $g'$, $\lambda$ in 
eq.(\ref{eq1}).

\section{ Nambu-Goldstone theorem (NGT) in L$\sigma$M loop order}
\label{sec4}

\begin{figure}[t]
\parbox{\halftext}{
  \epsfxsize=6.5 cm
  \epsfysize=1.2 cm
  \centerline{\epsffile{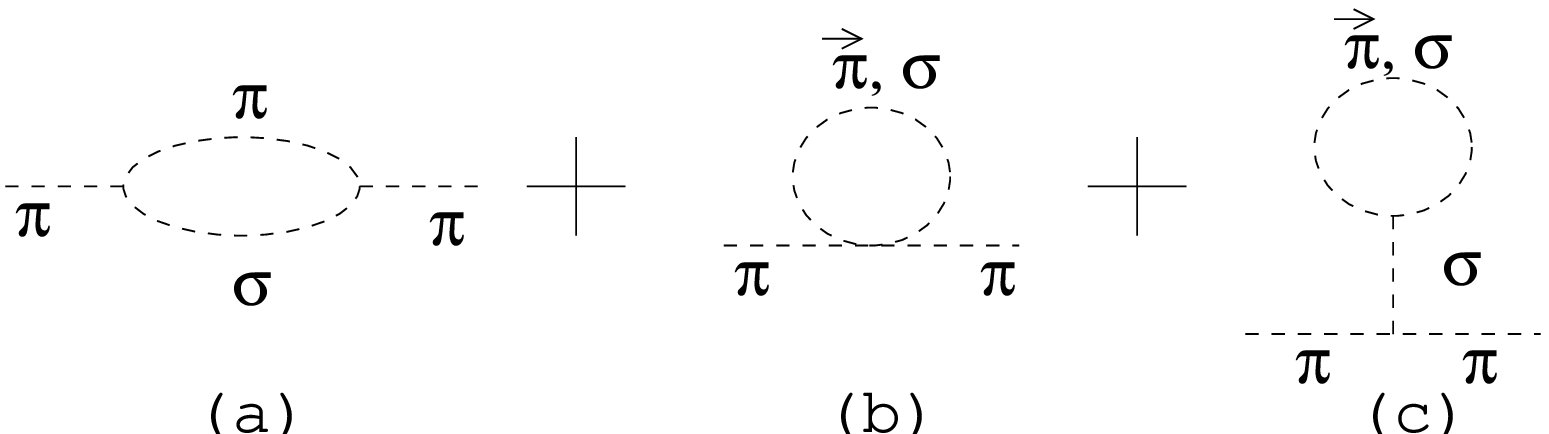}}
  \caption{Meson bubble(a) meson quartic (b) meson tadpole(c) graphs 
  for $m_\pi^2$.}
  \label{fig6}}
 \hspace{4mm}
 \parbox{\halftext}{
  \epsfxsize=6.5 cm
  \epsfysize=1.2 cm
  \centerline{\epsffile{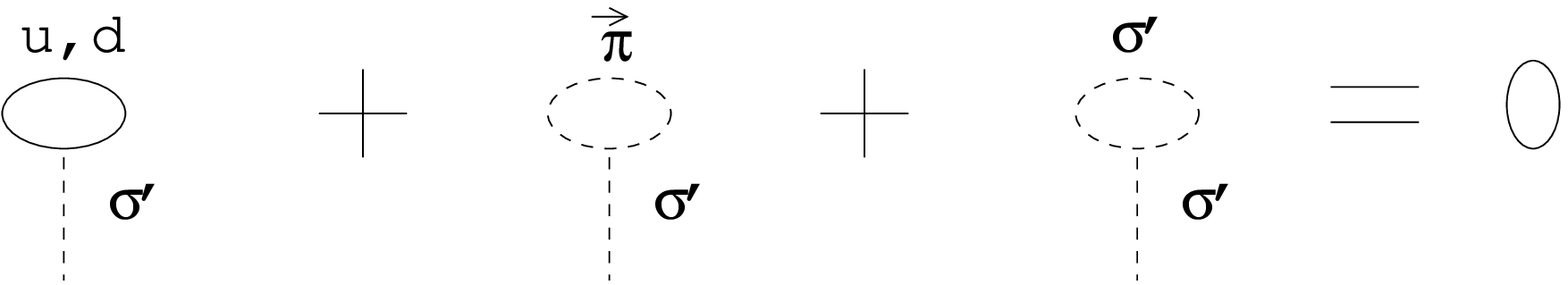}}
  \caption{Null tadpole sum for SU(2) L$\sigma$M.}
  \label{fig7}}
\end{figure}

Having dynamically generated the chiral 
pion and $\sigma$ as elementary, we must add to 
Fig.\ref{fig3} the five meson loops of Fig.\ref{fig6}. 
The first bubble graph in Fig.\ref{fig6} is log divergent, 
while the latter four quartic and tadpole graphs are quadratic divergent.

To proceed, first one uses a partial fraction identity to rewrite the 
log-divergent bubble graph as the difference of $\pi$ and $\sigma$ quadratic 
divergent integrals \cite{rf2,rf7}. Then the six meson loops (ml) of 
Fig.\ref{fig6} can be separated into three quadratic divergent $\pi$ and 
three quadratic divergent $\sigma$ integrals\cite{rf7}:
\begin{eqnarray}
 m^2_{\pi,ml} =(-2\lambda+5\lambda-3\lambda)i\int (p^2-m^2_\pi)^{-1}
  \bar{\mbf d}^4 p
  +(2\lambda+\lambda-3\lambda)i\int (p^2-m^2_\sigma)^{-1} \bar{\mbf d}^4 p.
  \ \ \ \ \ \ \ \ \ \ 
\label{eq17}
\end{eqnarray}
Adding eq.(\ref{eq17}) to eq.(\ref{eq12}), the total $m_\pi^2$
in the CL is in loop order
\begin{eqnarray}
 m_\pi^2=m_{\pi,ql}^2+ m_{\pi,\pi l}^2+ m_{\pi,\sigma l}^2 = 0 + 0 + 0 =0.
\label{eq18}
\end{eqnarray}
Moreover, eq.(\ref{eq18}) is chirally regularized and renormalized because 
the tadpole graphs of Figs.\ref{fig3}, \ref{fig6}c are already counterterm 
masses acting as subtraction constants.

A second aspect of the chiral pion concerns the pion charge radius $r_\pi$ 
in the CL. First one computes the pion form factor $F_{\pi,ql}(q^2)$ due to 
quark loops (ql) 
and then differentiates it with respect to $q^2$ at $q^2$ = 0 to find 
$r_{\pi, ql}^2$ as
\begin{eqnarray}
    r_{\pi, ql}^2 &=& \left.\frac{6dF_{\pi, ql}(q^2)}{dq^2}\right|_{q^2=0}
   = 8iN_cg^2\int_0^1dx6x(1-x)\int
     (p^2-m_q^2)^{-3} \bar{\mbf d}^4 p \nonumber\\
   &=& 8iN_c (4\pi^2/N_c)\cdot
       (-i\pi^2/2m_q^2 16\pi^4)= 1/m_q^2.
\label{eq19}
\end{eqnarray}

Although $r_\pi$ was originally expressed as 
$\sqrt{N_c}/2\pi f_\pi$\cite{rf8,rf7}, 
we prefer the result (\ref{eq19}) or $r_\pi = 1/m_q$, 
as it requires the tightly bound $q\bar{q}$ pion to have the two quarks 
{\it fused} in the CL. 
Later we will show that $N_c$ = 3; $m_q \approx$ 325 MeV in the CL gives 
$r_\pi =1/m_q \approx$ 0.6 fm. The observed $r_\pi$ is \cite{rf9} 
(0.63$\pm$0.01) fm. 
The alternative ChPT requires $r_\pi \propto L_9$, a low energy 
constant (LEC)! However VMD successfully predicts
\begin{eqnarray}
 r_\pi^{VMD}=\sqrt{6}/m_\rho\approx 0.63fm,
\label{eq20}
\end{eqnarray}
not only accurate but $r_\pi^{VMD}$ and $r_\pi^{L\sigma M}$ in (\ref{eq19}) 
and (\ref{eq20}) are clearly related\cite{rf7}.

\section {Lee null tadpole sum in SU(2) L$\sigma$M finding $N_c$ = 3}
\label{sec5}

To characterize the true DSB (not the false SSB) vacuum, B.~Lee \cite{rf10} 
requires the {\it sum} of loop-order tadpoles to vanish, see {\it eg.} 
our Fig.\ref{fig7}. 
This tadpole sum is\cite{rf3}
\begin{eqnarray}
 <\sigma '> =0 =-i8N_c gm_q \int(p^2-m_q^2)^{-1} \bar{\mbf d}^4 p 
   + 3ig'\int (p^2-m_\sigma^2)^{-1} \bar{\mbf d}^4 p.
\label{eq21}
\end{eqnarray}
Replacing $g$ by $m_q/f_\pi$, $g'$ by $m^2_\sigma/2f_\pi$ and scaling 
the quadratic divergent $q$(or $\sigma$) loop integrals by $m^2_q$ 
(or $m^2_\sigma$), eq.(\ref{eq21}) requires \cite{rf3}
(neglecting the pion tadpole)
\begin{eqnarray}
 N_c(2m_q)^4=3m_\sigma^4.
\label{eq22}
\end{eqnarray}
But we know from eq.(\ref{eq10}) that $2m_q = m_\sigma$, so the loop-order 
SU(2) L$\sigma$M result 
(\ref{eq22}) in turn {\it predicts} $N_c = 3$, a satisfying result ! 
Then the dynamically generated SU(2) loop-order L$\sigma$M 
in \S \ref{sec3} also 
predicts in the CL\cite{rf3} $m_q\approx$ 325 MeV, $m_\sigma\approx$ 650 MeV
and $g = 2\pi/\sqrt{3}$=3.6276, 
$g' = 2gm_q\approx$2.36 GeV, $\lambda=8\pi^2/3\approx$26.3.

\section{ Chiral $s$-wave cancellations in L$\sigma$M}
\label{sec6}
Away from the CL, the tree-order L$\sigma$M requires the cubic meson 
coupling to be
\begin{eqnarray}
 g_{\sigma\pi\pi}=(m_\sigma^2-m_\pi^2)/2f_\pi =\lambda f_\pi.
\label{eq23}
\end{eqnarray}
But at threshold $s = m^2_\pi$ , so the net $\pi\pi$ amplitude then vanishes
using (\ref{eq23}):
\begin{eqnarray}
 M_{\pi\pi}=M_{\pi\pi}^{contact} + M_{\pi\pi}^{\sigma pole} \rightarrow \lambda 
  + 2g_{\sigma\pi\pi}^2 (m_\pi^2 - m_\sigma^2)^{-1} =0.
\label{eq24}
\end{eqnarray}
In effect, the contact $\lambda$ `` chirally eats" the $\sigma$ pole 
at the $\pi\pi$ threshold at tree level. Then $\sigma$ poles from the 
cross channels predict a L$\sigma$M Weinberg PCAC form \cite{rf11,rf12}  
\begin{eqnarray}
M_{\pi\pi}^{abcd} &=& A\delta^{ab}\delta^{cd} + B\delta^{ac}\delta^{bd}+C \delta^{ad} \delta^{bc}, 
   \nonumber\\
A^{L\sigma M} &=& -2\lambda 
         \left[ 1- \frac{2\lambda f_\pi^2}{m_\sigma^2-s}\right]
              = \left(\frac{m_\sigma^2-m_\pi^2}{m_\sigma^2-s}\right)
                \left(\frac{s-m_\pi^2}{f_\pi^2}\right).
\label{eq25}
\end{eqnarray}
So the I=0 $s$-channel amplitude $3A+B+C$ at threshold predicts 
a 23\% enhancement of the Weinberg $s$-wave I=0 scattering length 
at $s= 4m^2_\pi$, $t=u= 0$ for $m_\sigma\approx$650 MeV with 
$\epsilon = m^2_\pi/m^2_\sigma\approx$0.045 and\cite{rf12}
(using only eq.(\ref{eq25}))
\begin{eqnarray}
 a_{\pi\pi}^{(0)}|_{L\sigma M}=\left(\frac{7+\epsilon}{1-4\epsilon}\right)
 \frac{m_\pi}{32\pi f_\pi^2} \approx (1.23) \frac{7m_\pi}{32\pi f_\pi^2} \approx 0.20m_\pi^{-1}.
\label{eq26}
\end{eqnarray}
For a $\sigma$(550) and $\epsilon\approx$0.063 this L$\sigma$M 
scattering length (\ref{eq26}) increases to 0.22$m_\pi^{-1}$. 
Compare this simple L$\sigma$M tree order result (\ref{eq26}) 
with the analogue ChPT 0.22 $m_\pi^{-1}$ scattering length requiring 
a 2-loop calculation involving about 100 LECs ! These $\pi\pi$ scattering 
length problems should be sorted out soon
by R. Kami\'nski, {\it et. al.} \cite{rf13}.

In L$\sigma$M loop order the analog cancellation is due to 
a Dirac matrix {\it identity}\cite{rf14}:
\begin{eqnarray}
 (\gamma\cdot p-m)^{-1} 2m\gamma_5 (\gamma\cdot p-m)^{-1} = 
 -\gamma_5 (\gamma\cdot p-m)^{-1}
 - (\gamma\cdot p-m)^{-1} \gamma_5.
\label{eq27}
\end{eqnarray}
At a soft pion momentum, eq.(\ref{eq27}) requires a $\sigma$ meson 
to be ``eaten" via a quark box-quark triangle cancellation for 
$a_1 \rightarrow \pi(\pi\pi)$ $s$ wave,  $\gamma\gamma\rightarrow2\pi^0$, 
$\pi^-p \rightarrow \pi\pi n$ as suggested in each case by low energy 
data\cite{rf14,rf15}. Also a soft pion scalar kappa $\kappa$(800-900) 
is ``eaten'' in $K^-p \rightarrow K^-\pi^+n$ peripheral 
scattering\cite{rf15}.

As for resonant $\pi\pi$ phase shifts, the needed $s$-wave chiral 
cancellation of eq.(\ref{eq24}) must have an analog $s$-wave $\pi\pi$ 
phase shift effect. S.~Ishida et. al.\cite{rf16} suggest 
this chiral cancellation corresponds to subtracting a background phase 
$-\delta_{BG}= p_\pi^{CM} r_c$, where $r_c$ is near the observed pion 
charge radius of 0.63 fm in eq.(\ref{eq19}) for the L$\sigma$M. 
Then CERN-Munich modified $\pi\pi$ phase shifts appear to approach 90 deg., 
corresponding to a resonant $\sigma$(550-600). 
This resonant $\sigma\rightarrow \pi^+\pi^-$ width is \cite{rf16}
\begin{eqnarray}
 \Gamma_R(s)=p_\pi^{CM}(8\pi s)^{-1}|g_R F(s)|^2 \approx {\rm 340 MeV}
 \ {\rm at} 
 \sqrt{s_R}\approx {\rm 600 MeV}, g_R \approx {\rm 3.6 GeV} \ \ \ \ \ \ \
\label{eq28}
\end{eqnarray}
for $p_\pi^{CM}=\sqrt{s/4-m^2_\pi}\approx$ 260 MeV. Then the total 
$\sigma \pi^+ \pi^-$, $\sigma \pi^0 \pi^0$ 
width is $\Gamma_{\sigma\pi\pi}\approx 3/2 \cdot$ 340 MeV = 510 MeV. 
Also this $g_R$ is double the L$\sigma$M coupling (\ref{eq23})\cite{rf12}: 
\begin{eqnarray}
 g_R \rightarrow 2g_{\sigma\pi\pi} =(m_\sigma^2 -m_\pi^2)/f_\pi 
 \approx 3.67 {\rm GeV},
\label{eq29}
\end{eqnarray}
close to $g_R\approx$ 3.6 GeV in eq.(\ref{eq28}). 
Furthermore the L$\sigma$M total decay width is\cite{rf17}
\begin{eqnarray}
 \Gamma_{\sigma\pi\pi}^{L\sigma M}=(3/2)\cdot(p_\pi^{CM}/8\pi)\cdot
   (|2g_{\sigma\pi\pi}|^2/m_\sigma^2)\approx {\rm 580 MeV},
 {\rm for}\  m_\sigma^R\approx {\rm 600 MeV}.\ \ \
\label{eq30}
\end{eqnarray}

\section{ VMD and the L$\sigma$M}
\label{sec7}

\begin{figure}[t]
  \epsfysize=1.1 cm
 \centerline{\epsffile{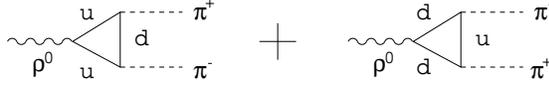}}
 \caption{Quark triangle graphs contributing to $\rho^0\rightarrow\pi\pi$.}
  \label{fig8}
\end{figure}

Given the implicit LDGE (\ref{eq3}) UV cutoff $\Lambda\approx$ 750 MeV, 
the $\rho$(770) can be taken as an external field (bound state $\bar{q}q$ 
vector meson). Accordingly the quark loop graphs of Fig.\ref{fig8} generate 
the loop order $\rho\pi\pi$ coupling \cite{rf2,rf3}
\begin{eqnarray}
 g_{\rho\pi\pi}=g_\rho[-i4N_cg^2\int(p^2-m_q^2)^{-2}
  \bar{\mbf d}^4 p]
 =g_\rho
\label{eq31}
\end{eqnarray}
via the LDGE (\ref{eq3}). While the individual udu and dud quark graphs of 
Figs.\ref{fig8} are both linearly divergent, when added together with vertices 
$g_{\rho^0 uu} = -g_{\rho^0 dd}$, the net $g_{\rho\pi\pi}$ loop in 
Fig.\ref{fig8} is log divergent. Equation(\ref{eq31}) is Sakurai's VMD 
universality condition. Also a $\pi^+ \sigma \pi^+$ meson loop added to 
the quark loops of Fig.\ref{fig8} gives\cite{rf7}
\begin{eqnarray}
 g_{\rho\pi\pi}=g_{\rho}+g_{\rho\pi\pi}/6\ 
 {\rm or}\ g_{\rho\pi\pi}/g_\rho = 6/5.
\label{eq32}
\end{eqnarray}

If one first gauges the L$\sigma$M lagrangian, the inverted squared gauge coupling is related to 
the $q^2 = 0$ polarization amplitude as \cite{rf3}
\begin{eqnarray}
 (g_\rho^{-2}) =\pi(0,m_q^2)
 =-8iN_c/6\cdot\int (p^2-m_q^2)^{-2} \bar{\mbf d}^4 p
 =(3g^2)^{-1}
\label{eq33}
\end{eqnarray}
by virtue of the LDGE(\ref{eq3}). But since we know $g=2\pi/\sqrt{3}$, 
eq.(\ref{eq33}) requires $g_\rho = \sqrt{3}g = 2\pi$, reasonably near 
the observed values $g_{\rho \pi\pi}\approx$ 6.05 
and $g_\rho\approx$ 5.03.

The chiral KSRF relation for the $\rho$ mass\cite{rf18} 
$m^2_\rho= 2g_{\rho\pi\pi}g_\rho f^2_\pi$ coupled with this L$\sigma$M 
implies $m^2_\rho = 2(2 \pi)^2 f_\pi^2  5/6\approx$ (754 MeV)$^2$, close to 
the observed $\rho$ mass.

\section{Scalar nonet and the IMF}
\label{sec8}
Although the SU(3) L$\sigma$M is more than a simple extension of the SU(2) 
L$\sigma$M\cite{rf19} (due to scalar and pseudoscalar mixing angles), 
one obtains kinematical insight to the SU(3) L$\sigma$M masses by working 
in the IMF because then the important dynamical tadpoles are substantially 
suppressed\cite{rf20} using IMF squared masses \cite{rf21}. 

Specifically, SU(6) equal splitting laws (ESLs) suggest 
\begin{eqnarray}
m_\kappa^2{\rm (806)} - m_\sigma^2{\rm (650)} 
= K^2{\rm (496)} - \pi^2{\rm (138)}.
\label{eq34}
\end{eqnarray}
Since the nonstrange $\eta$ mass is 759 MeV
(average of $\eta$(547), $\eta'$(958)), ESLs say\cite{rf21}
\begin{eqnarray}
 \sigma_{NS}^2 {\rm (650)} -\pi^2 {\rm (138)} &=& 
   \kappa^2{\rm (806)}-K^2{\rm (496)}
 =a_0^2{\rm (990)}-\eta_{NS}^2{\rm (759)}, \label{eq35}\\
\kappa^2{\rm (802)}- \sigma_{NS}^2 {\rm (650)}&=&
 \sigma_{S}^2 {\rm (930)}-\kappa^2{\rm (802)}.
\label{eq36}
\end{eqnarray}

Another approach is the dynamical NJL pattern \cite{rf19} away from the CL:
\begin{eqnarray}
 \sigma^2{\rm (680)}=4\hat{m}^2,\ \ \ \  \kappa^2{\rm (820)}=4m_s \hat{m}, 
 \ \ \ \ \sigma_S^2{\rm (950)}=4m_s^2,
\label{eq37}
\end{eqnarray}
for $m_s/\hat{m}\approx$ 1.45 and $\hat{m}\approx$ 340 MeV, 
the nonstrange constituent quark mass away from the CL. 
From eqs.(\ref{eq34}-\ref{eq37}), one may infer a scalar nonet pattern
 $\sigma_{NS}{\rm (650-680)}$, $\kappa{\rm (800-820)}$, 
$f_0{\rm (980)}$, $a_0{\rm (990)}$,
close to phase shift considerations \cite{rf16}.

It is interesting that the early $qq\bar{q}\bar{q}$ bag model suggested 
a similar scalar nonet pattern\cite{rf23} $\sigma{\rm (690)}$, 
$\kappa{\rm (880 ?)}$, $f_0{\rm (980)}$, $a_0{\rm (984)}$,
Moreover a unitarized coupled channel meson analysis\cite{rf24} 
obtains a scalar nonet $\sigma{\rm (500)}$, 
$\kappa{\rm (730)}$, $a_0{\rm (970)}$, $f_0{\rm (990)}$.

\section{L$\sigma$M as infrared limit of nonperturbative QCD}
\label{sec9}

We suggest five links between the L$\sigma$M and the infrared 
limit of QCD.
\begin{enumerate}
\renewcommand{\labelenumi}{\roman{enumi})}

\item Quark mass: the L$\sigma$M has $m_q=f_\pi\frac{2\pi}{\sqrt{3}}\approx$ 
325 MeV, while QCD has\cite{rf25}
$m_{dyn}=(\frac{4\pi\alpha_s}{3}<-\bar{\Psi}\Psi>_{\rm 1GeV})^{1/3}$
$\approx {\rm 320 MeV}$ at a 1 GeV near-infrared cutoff.

\item Quark condensate: the L$\sigma$M condensate is at infrared cutoff
$m_q$\cite{rf26}:\\
$$
<-\bar{\Psi}\Psi>_{m_q}=i4N_c m_q\int\frac{\bar{\mbf d}^4 p}{p^2-m_q^2}
=\frac{3m_q^2}{4\pi^2}\left[\frac{\Lambda^2}{m_q^2}-ln\left(\frac{\Lambda^2}{m_q^2}+1\right)\right]
\approx ({\rm 209 MeV})^3,
$$
while the condensate in QCD is
$<-\bar{\Psi}\Psi>_{m_q}=3m_{dyn}^3/\pi^3\approx ({\rm 215 MeV})^3.$

\item Frozen coupling strength: the L$\sigma$M coupling is for 
$g=2\pi/\sqrt{3}$ or
$ \alpha_{L\sigma M} = \frac{g^2}{4\pi} =\frac{\pi}{3}, $
while in QCD $\alpha_s=\frac{\pi}{4}$ at infrared freezeout\cite{rf27} leads to
 $\alpha_s^{eff}=(4/3)\alpha_s=\pi/3$.

\item $\sigma$ mass: 
the L$\sigma$M requires $m_\sigma=2m_q$, while the QCD condensate 
gives\cite{rf28}
$m_{dyn}=\frac{g_{\sigma qq}}{m_\sigma^2}<-\bar{\Psi}\Psi>_{m_\sigma}$
for $\alpha_s(m_\sigma)\approx\pi /4$, or
$m_\sigma^2/m_{dyn}^2=\pi/\alpha_s(m_\sigma^2)\approx 4.$

\item Chiral restoration temperature $T_c$: 
the L$\sigma$M requires\cite{rf29} $T_c=2f_\pi\approx 180$ MeV, 
while QCD computer lattice 
simulations find\cite{rf30} $T_c=$(150$\pm$30) MeV.

\end{enumerate}

\section{Conclusion}
\label{sec10}
In \S \ref{sec2}, \ref{sec3}, the SU(2) L$\sigma$M lagrangian 
was dynamically generated in all (chiral) regularization schemes, 
via loop gap equations, 
predicting the NJL $\sigma$ mass $m_\sigma$ = $2m_q$ along with meson-quark 
coupling $g= 2\pi/\sqrt{N_c}$. Then the three-and four-point quark loops were 
shown to ``shrink" to tree graphs, giving the meson cubic and quartic 
couplings $g' =m_\sigma^2 /2f_\pi$, $\lambda = 8 \pi^2/N_c$. 
Next in \S \ref{sec4}, the Nambu-Goldstone theorem (NGT) was shown to hold 
in L$\sigma$M loop order with the pion charge radius 
$r_\pi=1/m_q$. In \S \ref{sec5}, the SU(2) 
L$\sigma$M requires color number $N_c$=3 in loop order, 
then predicting $m_q\approx$ 325 MeV, 
$m_\sigma\approx$ 650 MeV, $g\approx$ 3.63, $\lambda\approx$ 26, 
$r_\pi\approx$ 0.6 fm in the CL.

In \S \ref{sec6}, we considered L$\sigma$M chiral cancellations, both in 
tree and in loop order. 
Next, in \S \ref{sec7}, Sakurai's vector meson dominance (VMD) empirically 
accurate scheme follows from the L$\sigma$M, the latter further predicting 
$g_{\rho\pi\pi} = 2\pi$ and $g_{\rho\pi\pi}/g_\rho$= 6/5 along with the KSRF 
relation. Then in \S \ref{sec8}, a global extension of the SU(2) L$\sigma$M 
to SU(3) was found in the infinite momentum frame (IMF). Finally in 
\S \ref{sec9}, we suggested that the L$\sigma$M is the infrared limit of 
nonperturbative QCD.

\end{document}